\documentclass[preprint2]{aastex}

\shorttitle{Raman-scattered \ion{O}{6} in Sanduleak's Star}
\shortauthors{Heo et al.}

\begin{document}


\title{A Profile Analysis of Raman-scattered \ion{O}{6} Bands at 6825 \AA\ and 
7082 \AA\ in Sanduleak's Star\footnote{This paper includes data gathered with the 6.5 meter Magellan Telescopes located at Las Campanas Observatory, Chile.}}


\author{Jeong-Eun Heo\altaffilmark{1,2}, Rodolfo Angeloni\altaffilmark{2}, Francesco Di Mille\altaffilmark{3}, Tali Palma\altaffilmark{4}, and Hee-Won Lee\altaffilmark{1}}
\email{jeung6145@gmail.com}

\altaffiltext{1}{Department of Physics and Astronomy, Sejong University, Seoul, Korea}
\altaffiltext{2}{Gemini Observatory, Casilla 603, La Serena, Chile}
\altaffiltext{3}{Las Campanas Observatory, Carnegie Observatories, Casilla 601, La Serena, Chile}
\altaffiltext{4}{Departamento de Ciencias F\'isicas, Universidad Andr\'es Bello, Fernández Concha 700, Las Condes, Santiago, Chile}



\begin{abstract}
We present a detailed modeling of the two broad bands observed at 6825~\AA\ and 7082~\AA\ in Sanduleak's star, a controversial object in the Large Magellanic Cloud.
These bands are known to originate from Raman-scattering of \ion{O}{6}~$\lambda\lambda$~1032 and 1038 photons with atomic hydrogen and are only observed in \textit{bona fide} symbiotic stars. 
Our high-resolution spectrum obtained with the \textit{Magellan Inamori Kyocera Echelle} (MIKE) spectrograph at the Magellan-Clay Telescope reveals, quite surprisingly, that the profiles of the two bands look very different: while the Raman 6825~\AA\ band shows a single broad profile with a redward extended bump, the Raman 7082~\AA\ band exhibits a distinct triple-peak profile. 
Our model suggests that the \ion{O}{6} emission nebula can be decomposed into a red, blue and central emission regions from an accretion disk, a bipolar outflow and a further compact, optically thick region. 
We also perform Monte Carlo simulations with the aim of fitting the observed flux ratio $F(6825)/F(7082) \sim 4.5$, which indicate that the neutral region in Sanduleak's star is characterized by the column density $N_{HI} \sim 1 \times 10^{23} {\rm\ cm^{-2}}$.
\end{abstract}



\section{Introduction}\label{sec:intro}
Sanduleak's star was discovered in 1977 as an emission line variable object in the Large Magellanic Cloud \citep{sanduleak77}. 
A few years later, \cite{allen80} advocated its nature of symbiotic star, noticing however that the optical continuum appeared weak and featureless, with no clear signature of any late-type giant -- which nonetheless is a necessary ingredient at the basis of the symbiotic phenomenon.
After almost forty years, the real nature of Sanduleak's star is still a matter of debate: despite no observations unambiguously confirming the presence of a late-type companion, in favor of its binary nature there is the bipolar precessing jet recently discovered by \cite{angeloni11}. As a matter of fact, bipolar outflows seem to be associated with the binarity of their central source \citep{sahai11}.
Furthermore, the symbiotic character of Sanduleak's star is supported by both a highly ionized emission line spectrum reminiscent of a dusty (hereafter D-) type symbiotic star \citep{munari02,belc00} and by the two intense Raman \ion{O}{6} bands at 6825, 7082~\AA, which so far have been convincingly detected only in \textit{bona fide} symbiotic stars. 

\textit{De facto}, these Raman-scattered \ion{O}{6} bands are so unique to the symbiotic phenomenon that their presence has been used as a sufficient criterion for classifying a star as symbiotic, even in those cases where the cool companion appears to be hiding \citep{belc00}.
These broad spectral bands are known in the astronomical literature at least since the 1940's \citep{joy45}, but their physical origin was explained only much later by \cite{schmid89}, who proposed that an
\ion{O}{6}~$\lambda$~1032 photon incident on a hydrogen atom in the ground $1s$ state may 
be Raman-scattered to become an optical photon with $\lambda$=6825~\AA, leaving the scattering 
hydrogen atom in the excited $2s$ state; an analogous process for 
\ion{O}{6}~$\lambda$~1038 photons produce the Raman band at 7082~\AA.

The resonance doublet \ion{O}{6}~$\lambda\lambda$ 1032 and 1038 arises from $S_{1/2}-P_{3/2, 1/2}$ transitions. As $P_{3/2}$ states have twice more sublevels than 
$P_{1/2}$ states, under optically thin conditions the \ion{O}{6}~$\lambda$~1032 flux, $F(1032)$, is expected to be twice stronger than the \ion{O}{6}~$\lambda$~1038 one, $F(1038)$.  However, when the emission region is optically thick, $F(1032)/F(1038)\approx 1$ because of thermalization effects (\citealp{kang08,schmid99}).
Since symbiotic nebulae are highly stratified \citep{allen87,luna05}, it is indeed quite common to observe UV resonance doublets with different flux ratios, i.e., emerging from regions with different optical depths (as in CI~Cyg, \citealp{miko06}). 

It is also important to emphasize that, 
while the \ion{O}{6}~$\lambda\lambda$~1032 and 1038 line profiles are dependent on the observer's line of sight (e.g., in EG~And, their broad components completely disappear when the white dwarf is at the superior conjunction -  \citealp{crowley08}), the Raman band profiles are not. 
Instead, they mainly reflect the relative kinematics between the \ion{H}{1} scattering region and the far-UV emission region:  therefore, the Raman band profiles observed in symbiotic binaries gives us the unique perspective of the mass transfer process as seen from the donor star \citep[e.g.,][]{schmid89,nussbaumer89}. 

Observationally, the Raman \ion{O}{6} bands at 6825~\AA\ and 7082~\AA\ tend to exhibit multiple-peaked profiles, in which the blue peak of Raman 6825~\AA\ is relatively more enhanced than the corresponding one of 7082~\AA\ \citep{harries96,schmid99}. 
\cite{lee07} performed profile analyses of Raman \ion{O}{6}~6825~\AA\ band observed in the two D-type symbiotic stars V1016~Cyg and HM~Sge to probe the kinematics of the emission region around the white dwarf. They attributed the observed asymmetric double-peak profiles to the \ion{O}{6} emission from an accretion disk: the peak separation corresponds to a speed of $\sim 50{\rm\ km\ s^{-1}}$, implying an emitting region of $\sim 1$~AU.

\citet{heo15} further noticed that in V1016 Cyg the blue peak of the 7082~\AA\ band is relatively weaker than the 6825~\AA\ counterpart, when the two Raman bands are normalized to an equal red peak strength in the Doppler factor space. They developed a quantitative model in which an accretion stream around the white dwarf is responsible for the double-peak profiles, and where the difference in the profile shape is due to the  $F(1032)/F(1038)$ variation in the accretion stream.

Spectropolarimetric data of the same V1016~Cyg by \cite{schild96} revealed that the red wing of Raman \ion{O}{6} red peak is polarized in the direction perpendicular to the polarization exhibited by the main part of the bands. 
This polarization flip is consistent with the emission region made of two components, one moving parallel and the other one perpendicularly (i.e., an outflow-like structure) with respect to the accretion disk plane. 
Also \cite{schmid00} showed a similar polarization pattern in HM Sge. It is consistent with the presence of a bipolar outflow which was reported by several observations \citep{solf84,eyres95}.
Since the multiple-peak structure and the polarization flip in the Raman \ion{O}{6} bands
can be intuitively explained by a model with an accretion disk and a bipolar outflow,
we then apply and test this theoretical scenario on Sanduleak's star. \\

In what follows, we present a detailed analysis of the Raman \ion{O}{6} bands in Sanduleak's star.
In \S~\ref{sec:obs} we present archival {\it Far Ultraviolet Spectroscopic Explorer} (FUSE) and proprietary \textit{Magellan Inamori Kyocera Echelle} (MIKE) data with a morphological description of the observed profiles of both the far-UV \ion{O}{6} doublet (\S~\ref{sec:fuse}) and the optical Raman \ion{O}{6} bands (\S~\ref{sec:mike}). Our profile decomposition technique, along with our physical interpretation that reads these profiles in terms of the relative kinematics between the \ion{O}{6} emission region and the \ion{H}{1} scattering region, is fully described in \S~\ref{sec:profile}.
In \S~\ref{sec:sim} we show the results of Monte-Carlo simulations that help constraining the physical conditions (e.g., column density and spatial extent) of the circumstellar \ion{H}{1} region. Discussion and concluding remarks follow in \S~\ref{sec:dis}.

\section{Observations and Profile Comparisons}\label{sec:obs}

Symbiotic stars are composite interacting binaries: at the origin of their complex photometric and spectroscopic variability lies a series of astrophysical phenomena (accretion processes, stellar pulsations and orbital modulations, just to quote a few) that act on time scales ranging from minutes to centuries \citep{munari12}. Any multi-wavelength study aiming at a self-consistent modeling of a symbiotic system should thus take into account all the caveats arising when combining observations taken over different spectral ranges at different epochs.

Specifically to the analysis of Raman-scattering processes, the ideal case would be that of having virtually simultaneous observations of both the far-UV \ion{O}{6} resonance doublet and the optical Raman \ion{O}{6} bands. 
\cite{birriel00} performed near simultaneous far-UV and optical observations to confirm the identification of Raman O VI bands and derive the Raman scattering efficiencies for 9 symbiotic systems \citep[see also][]{birriel98}.
Even when a high-resolution optical spectrum is at hand, however, far-UV data are as precious as rare. Luckily enough, a search in the Mikulski Archive for Space Telescopes (MAST) returned a publicly available UV spectrum of Sanduleak's star, which we publish here for the first time. 

\subsection{{\it FUSE} Spectrum of \ion{O}{6}~$\lambda\lambda$~1032, 1038}\label{sec:fuse}

The {\it Far Ultraviolet Spectroscopic Explorer} (FUSE) was launched in 1999 and operated until the failure of the pointing system on the satellite, on October 18, 2007. At the conclusion of the mission, the FUSE data archive was moved to MAST, where all the data were re-processed and archived with the final version of the CalFUSE calibration pipeline software package (v 3.2, \citealp{dixon07}) and made available to the whole community. 

Through Program E950, FUSE pointed at Sanduleak's star for a total of $\sim$22,000 s on August 14, 2004. The science data were collected with the largest apertures on the focal plane assembly (LWRS - 30 arcsec$^2$) in TTAG photon collecting mode. They cover the entire wavelength region 905-1187 \AA, with a nominal spectral resolution of $\Delta v\approx 20 {\rm\ km\ s^{-1}}$ \citep{moos00,sahnow00}, and are placed by the same CalFUSE on a heliocentric velocity scale. 

\begin{figure}[!t]
\epsscale{1.}
\plotone{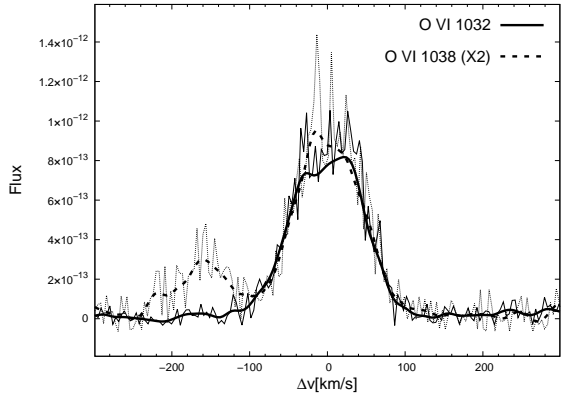}
\caption {FUSE spectrum of Sanduleak's star around \ion{O}{6}$\lambda\lambda$1032 and 1038 in the Doppler factor space.
The solid and dotted lines are for \ion{O}{6}~$\lambda$~1032, and \ion{O}{6}~$\lambda$~1038 (multiplied by 2), respectively. The thick lines show the profiles convolved with the instrumental profile (see the text). There exists a clear small excess at $\sim 20 {\rm\ km\ s^{-1}}$ in the \ion{O}{6}~$\lambda$~1038 line. The emission at $\sim -150 {\rm\ km\ s^{-1}}$ blueward of \ion{O}{6} $\lambda$ 1038 is \ion{C}{2}~$\lambda$~1037. }
\label{FUSE}
\end{figure}

The \ion{O}{6}~$\lambda$ 1032 feature appears among the most intense emission lines present in the far-UV spectrum of Sanduleak's star, second to (in decreasing order) Ly$\beta$, \ion{C}{3}~$\lambda$ 977 and Ly~$\gamma$. In the present work we analyse and discuss only the profiles of the \ion{O}{6}~$\lambda\lambda$~1032 and 1038 resonance doublet (Fig.~\ref{FUSE}), postponing to a forthcoming paper the  modeling of the entire FUSE emission line spectrum.

Following \cite{nichols09}, in Fig.~\ref{FUSE} we overplot the two \ion{O}{6} line profiles: 
the horizontal axis sets the Doppler factor space, whose origin has been chosen at the (respective) line central wavelength in vacuum \citep{moore79}:
\begin{equation}
\lambda_{1032}=1031.928\ {\rm\ \AA},\ \lambda_{1038}=1037.618\ {\rm\ \AA},
\end{equation}
and where \ion{O}{6}~$\lambda$~1038 has been multiplied by two in order to focus on the
profile comparison. 
Based on Fig. 5 of \cite{wood02} we take the instrumental profile, or line-spread function, as a single Gaussian having the FWHM of $ 20 {\rm\ km s^{-1}}$. The convolved profiles are plotted by the thick lines in Fig.~\ref{FUSE}.
It is immediately evident that the two profiles coincide with each other, with the remarkable exception of a small excess at a few tens $\rm\ km\ s^{-1}$ in the \ion{O}{6}~$\lambda$~1038 line. This excess implies that locally $F(1032)/F(1038)$ is less than 2: as remembered in \S~\ref{sec:intro} about far-UV resonance doublets in symbiotic systems, this evidence hints toward the presence in Sanduleak's star of an \ion{O}{6} optically thick region that is approaching the observer with a velocity of $\sim 20 {\rm\ km\ s^{-1}}$. A description of this emitting component in the framework of our scenario is discussed in \S~\ref{sec:thick}.

\subsection{\textit{MIKE} Spectrum of Raman \ion{O}{6} Bands}\label{sec:mike}

\begin{figure*}[!t]
\centering
\epsscale{2.}
\plottwo{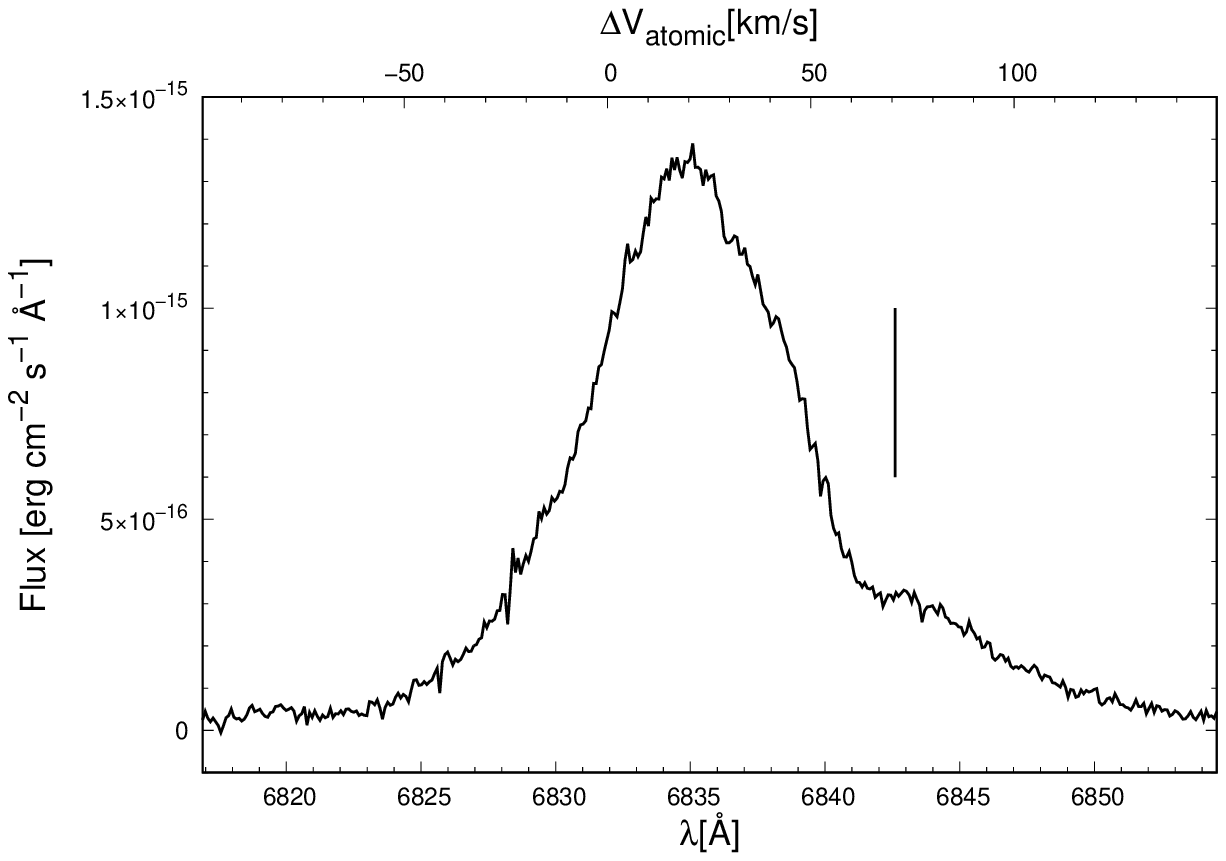}{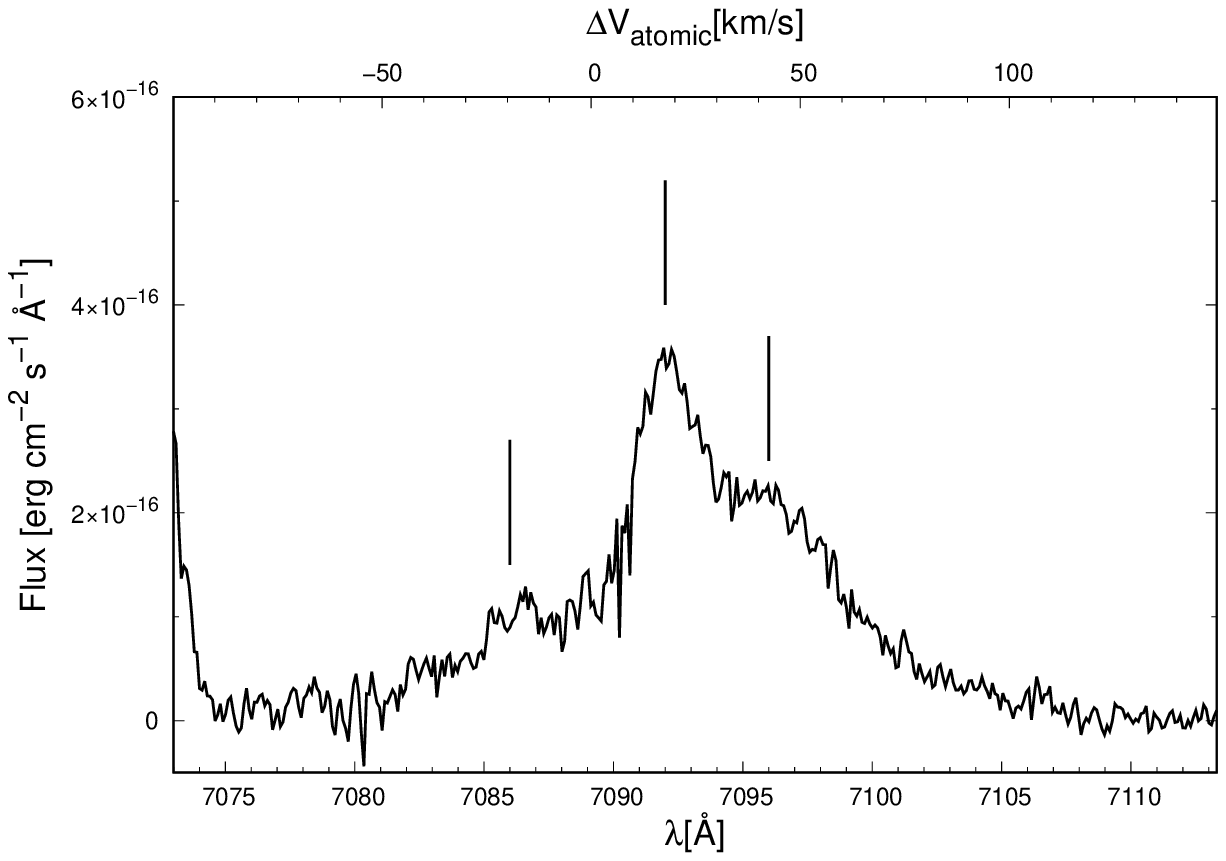}
\caption{The Raman-scattered \ion{O}{6} bands at 6825~\AA\ (left) and 7082~\AA\ (right) in Sanduleak's star. 
The upper horizontal axis corresponds to the Doppler factor space $\Delta V_{atomic}$, whose origin has been chosen at the \ion{O}{6} line central wavelength in vacuum. The vertical bar in the left panel indicates the red bump and the three vertical bars in the right panel show the three peaks.}
\label{MIKE}
\end{figure*}

Optical high-resolution spectroscopic observations of Sandulak's star were carried out on 2010 November 21 using the MIKE spectrograph at the 6.5 m  Magellan-Clay telescope, Las Campanas Observatory in Chile. The spectrograph delivers full wavelength coverage from about 4,900-10,000~\AA\ in its \textit{red} configuration: with the 0.7$\times$5 arcsec slit we used, the spectral resolution was R$\sim$32,000 \citep{bernstein03} . The scale of the CCD is about 7.5 pix/arcsec (0.13 arcsec/pix), therefore we binned the detector in the spectral and spatial direction to maximize S/N ratio. We took a series of 3$\times$900 sec exposures. Data reduction was performed under \texttt{IRAF} following the standard procedure. 

In Fig.~\ref{MIKE}, we display the Sanduleak's star  Raman-scattered \ion{O}{6}~$\lambda\lambda$ bands at 6825~\AA\ and 7082~\AA. It is noticeable that the two bands have quite different profiles: while the Raman 6825~\AA\ band exhibits a single-peak profile with a redward extended bump, the Raman 7082~\AA\ band reveals a distinct triple-peak profile. 
In order to make a quantitative comparison of the two optical profiles, we convert the observed wavelength to the \ion{O}{6} parent Doppler factor as follows: adopting the refractive index of air $n_{air}=1.0002763$ \citep{ciddor96}, the air wavelength $\lambda_{obs}$ is transformed into the vacuum wavelength $\lambda_{vac}$ by
\begin{equation}
\lambda_{obs}n_{air}=\lambda_{vac}.
\end{equation}
 
We then introduce $\lambda_f$, the vacuum wavelength corrected for the systemic velocity $v_{sys}$, as
\begin{equation}
\lambda_f={{\lambda_{vac}} \over {1+{v_{sys} \over c}}}.
\end{equation}
Using the energy conservation principle, we find the corresponding far-UV wavelength of incident \ion{O}{6} photon
\begin{equation}
\lambda_{UV}={\lambda_{f}\lambda_{Ly\alpha}\over{\lambda_{f}+\lambda_{Ly\alpha}}},
\end{equation}
where $\lambda_{Ly\alpha}=1215.668\ {\rm \AA}$. 

The \ion{O}{6} central wavelengths of Eq. (1) are then used to obtain the Doppler factors $\Delta V_{atomic}$:  
 \begin{equation}
{\Delta V_{atomic} \over c}={{\lambda_{UV}-\lambda_{i}}\over{\lambda_{i}}},\ i=1032, 1038.
\end{equation}

\section{Profile Decomposition}\label{sec:profile}

\begin{figure}[!h]
\epsscale{.9}
\plotone{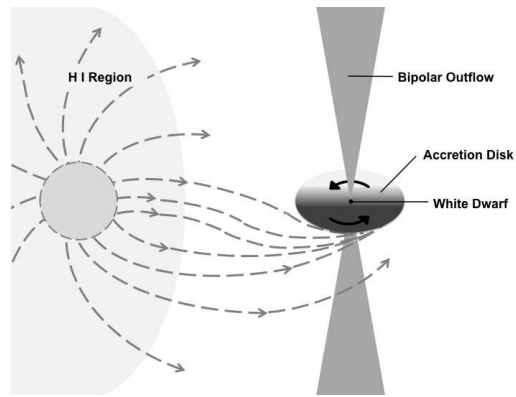}
\caption{Schematic model of Sanduleak's star. The \ion{O}{6} emission region
is assumed to consist of an accretion disk, a bipolar outflow and a further optically thick compact component. See text for details.}
\label{model}
\end{figure}

The presence in Sanduleak's star of a highly-collimated jet \citep{angeloni11} and of Raman \ion{O}{6} bands exhibiting multiple-peak profiles (Fig.~\ref{MIKE}) implies that a plausible model of its \ion{O}{6} emission nebula should at least foresee an accretion disk and a bipolar outflow. Moreover, the $F(1032)/F(1038)$ deviation from 2 discussed at the end of \S~\ref{sec:fuse} calls for the inclusion of an additional emitting region characterized by a high optical depth. Our schematic layout of Sanduleak's star is presented in Fig.~\ref{model}. 

 Based on the observed profiles of the Raman-scattered \ion{O}{6} bands presented in Fig.~\ref{MIKE}, we reconstruct in Fig.~\ref{input} the \ion{O}{6} far-UV resonance doublet by combining five emitting components: three arising from the accretion disk; one from the bipolar outflow; one from the optically thick nebula. For the sake of simplicity, each component is described by a Gaussian function characterized by a central velocity $\Delta V_{atomic}$, a FWHM $\Delta v$ and a peak value $f$, the latter normalized at the peak value of the 1032 line. 
 
Before the detailed discussion of each component with our corresponding physical interpretation, it is worth clarifying once again that the reconstructed \ion{O}{6} $\lambda\lambda$ 1032 and 1038~\AA\ line profiles are the ones seen by a hypothetical observer at rest with respect to the scattering region (that we assume stationary with respect to the donor star) and can perfectly be (in fact, almost always are) different from the actual observed profiles (Fig.~\ref{FUSE}), which are strongly dependent on the observer point of view. 

\subsection{Blue Emission Part (BEP) and Red Emission Part (REP) from the Accretion Disk}\label{sec:beprep}

As shown in the previous sections, the Raman \ion{O}{6}~7082~\AA\ band exhibits a quite distinct triple-peak structure that is not reproduced in the Raman \ion{O}{6}~6825~\AA\ band -- probably because severely overwhelmed by the other emission components. Since the central peak is considerably narrower than the blue and the red peaks, we explain these latter ones as originating from the accretion disk, and fit them with $\Delta v \sim 28{\rm\ km\ s^{-1}}$.
The association of the blue and red peaks with an accretion disk is supported by their peak separation ($\sim 70 {\rm\ km\ s^{-1}}$) which is comparable to Keplerian motions proposed for other D-type symbiotic stars \citep{lee07}. 

Since our interest is mainly focused on the kinematics of the \ion{O}{6} emission region with respect to the neutral hydrogen scattering region, we introduce the parameter $\Delta V_{O VI}$, and sets its zero-point at the average velocity of the blue and red peaks (shown by dotted lines in Fig.~\ref{input}). 

The Gaussian component corresponding to the blue peak has $\Delta V_{O VI}=-34{\rm\ km\ s^{-1}}$, and stems from the part of the accretion disk \textit{approaching the scattering region}. The Gaussian component corresponding to the red peak has $\Delta V_{O VI}=+34{\rm\ km\ s^{-1}}$, and it comes from the part of the accretion disk which is \textit{receding from the scattering region}. 
Hereafter, we refer to these subregions of the accretion disk as Blue Emission Part (BEP) and Red Emission Part (REP), respectively. 

By noticing that the red peak is conspicuously stronger than the blue peak, we can further deduce that the accretion flow must be quite asymmetric.
Since $F(1032)/F(1038)$ decreases from 2 to 1 as the optical depth of an emitting region increases, we naturally assign $F(1032)/F(1038)= 2$ to BEP and 1 to REP.
This general result is consistent with hydrodynamic simulations of a white dwarf and a mass losing giant in wide binary systems, which show how a stable accretion disk can be formed with size ranging from sub AU to tens of AU \citep[e.g.][]{huarte13,deval09, mastro98}, and also with 3D high-resolution hydrodynamic simulations of the recurrent nova RS Oph \citep{walder08}, which clearly show a density asymmetry in the wind accretion flow.

\subsection{Central Emission Part (CEP) in the Accretion Disk}\label{sec:cep}
Because the line profile from a disk in Keplerian motion has a non-zero flux at line center, just summing up the BEP and REP contributions fails to account for the emission from those parts of the accretion disk with null radial velocity component. 
Also, in order to explain why the Raman \ion{O}{6} 7082~\AA\ band exhibits a triple peak structure while the \ion{O}{6}~6825~\AA\ band does not, it is necessary to invoke the contribution in the \ion{O}{6} emission region of an optically thin gas 
characterized by $F(1032)/F(1038)=2$ and $\Delta V_{O VI}\sim 0$, i.e., at rest with respect to the scattering region. In analogy with the previous case, we name this component Central Emission Part (CEP), where with the word ``center'' we just identify that emitting component that fills the gap between REP and BEP in the Doppler factor space.

Because this optically thin component is quite significant in flux, the associated emitting region must be also quite extended: the considerable emission volume leads to a wide occupancy in velocity space, resulting thus in broad Gaussians. Our best fit is obtained for $\Delta V_{O VI}=6 {\rm\ km\ s^{-1}}$, $\Delta v \sim 42{\rm\ km\ s^{-1}}$ and when the \ion{O}{6}~1038 peak value is half the \ion{O}{6}~1032 one.

\subsection{Bipolar Outflow}\label{sec:bipolar}
The red bump feature visible in the \ion{O}{6}~6825~\AA\ band has a representative speed of $\Delta V_{atomic}= +71{\rm\ km\ s^{-1}}$, which corresponds to $\Delta V_{O VI}=+57{\rm\ km\ s^{-1}}$. No parallel feature is revealed in the \ion{O}{6} 7082~\AA\ band at the same velocity position, though. 

To explain such (lack of) evidence we suggest that this emitting component has to be associated with the bipolar outflow, that appears receding from the scattering region. The best fit is for $\Delta v \sim 67{\rm\ km\ s^{-1}}$ and $F(1032)/F(1038)=2$. In this way, the contribution of this component to the \ion{O}{6}~7082~\AA\ profile
is relatively smaller than to the \ion{O}{6}~6825~\AA\ profile, explaining why the red bump clearly appears in the latter, but is absent in the former profile.

\subsection{Optically Thick Compact Component}\label{sec:thick}
Even with the inclusion of an optically thin component (\S~\ref{sec:cep}), the sharp central peak of the Raman 7082~\AA\ band is poorly reproduced. A more satisfactory fit could be obtained including a fairly weak, narrow component at $\Delta V_{O VI}\sim 0$. By introducing an emitting component characterized by $\Delta V_{O VI}=+5 {\rm\ km\ s^{-1}}$,
$\Delta v \sim 11{\rm\ km\ s^{-1}}$ and $F(1032)/F(1038)= 1$ we sensibly improve the 7082~\AA\ fit without altering the 6825~\AA\ profile in a significant way. Also, we naturally take into account the evidence given by the \textit{FUSE} spectrum about the existence of an optically thick nebula in Sanduleak's star (\S~\ref{sec:fuse}).  
This Gaussian function is enveloped by the Gaussian function used in \S~\ref{sec:cep}: we can not exclude the possibility that the optically thick compact component is spatially part of the extended tenuous component. The association of this component with a specific emitting region is however not intuitive far from simple. One may speculate that a local density enhancement in the accretion stream, for example in the form of spiral structures, could represent a possible explanation, as also suggested by the numerical work of \cite{walder08}.

Summarizing, Table~\ref{tbl-1} lists the resulting parameters of each Gaussian, while Table~\ref{tbl-2} reports on the relative contribution of each component to the overall profiles. As it can be seen, the dominant contribution is made by the accretion disk, which takes up 70\%, whereas the bipolar outflow contributes about 20\% for Raman 6825 and 15\% for Raman 7082.
Although the optically thick component contribute only less than 10\%, the inclusion in the model of this region is essential for explaining the small excess revealed by the {\it FUSE} spectrum and the triple-peak profile which is only visible in the Raman 7082 band.

\begin{figure}
\plotone{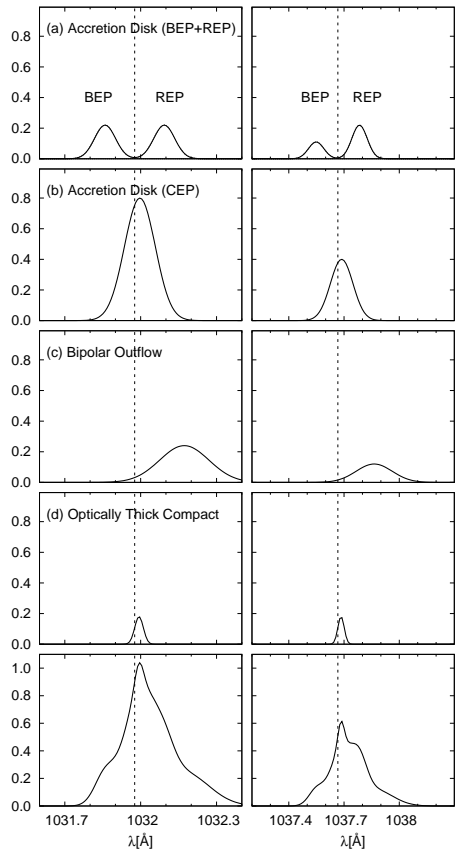}
\caption{Profile synthesis of the far-UV \ion{O}{6} $\lambda\lambda$ 1032, 1038 based on the observed Raman bands in Sanduleak's star. 
The \ion{O}{6} emitting nebula is reconstructed by summing up the contribution of three emitting regions from a) BEP, REP and b) CEP of the accretion disk, c) the bipolar outflow and d) the optically thick compact component. The bottom two panels show the reconstructed \ion{O}{6} $\lambda\lambda$~1032 (left) and 1038 (right) line profiles. The vertical dotted line in each plot represents the average velocity (taken as zero-point of $\Delta V_{O VI}$) of BEP and REP of the accretion disk. The parameters of the five Gaussian components are summarized in Table~\ref{tbl-1}. See also \S~\ref{sec:profile}. }
\label{input}
\end{figure}

\begin{table*}[!h]
\centering
\caption{Doppler factors, corresponding central wavelengths, FWHM ($\Delta v$), and peak values of the five Gaussian components discussed in \S~\ref{sec:profile} - see text for the details.}
\label{tbl-1}
\vskip5pt
 \begin{tabular}{cccccccc}
 \hline
 Emission Region & $\Delta V_{atomic}$ & $\Delta V_{O VI}$ & $\lambda_{1032}$ & $\lambda_{1038}$ & $\Delta v$ & $f_{1032}$ & $f_{1038}$ \\
 &($\rm\ km\ s^{-1}$) & ($\rm\ km\ s^{-1}$) & (\AA) & (\AA) & ($\rm\ km\ s^{-1}$)  & \\
 \hline
 Accretion Disk (BEP) & -20 & -34 & 1031.859 & 1037.549 & 28.3 & 0.22 & 0.11 \\
 Accretion Disk (REP) & 48 & 34 & 1032.093 & 1037.784 & 28.3  & 0.22 & 0.22 \\
 Accretion Disk (CEP) & 20 & 6 & 1032.997 & 1037.687 & 41.6 & 0.8 & 0.4  \\
 Bipolar Outflow & 71 & 57 & 1032.172 & 1037.864 & 66.6 & 0.24 & 0.12 \\
 Optically Thick Compact & 19 & 5 & 1032.993 & 1037.684 & 11.7 & 0.18 & 0.18 \\
 \hline

 \end{tabular}

\end{table*}

\begin{table}
\centering
\caption{Relative contribution of each Gaussian component to the overall \ion{O}{6} $\lambda\lambda$ 1032 \& 1038~\AA\ line profile (see also Fig.~\ref{input}).}
\label{tbl-2}
\vskip5pt
 \begin{tabular}{ccc}
 \hline
 Emission Region & 1032 & 1038 \\
 \hline
 Accretion Disk (BEP + REP) & 0.232 & 0.296 \\
 Accretion Disk (CEP) & 0.512 & 0.435 \\
 Bipolar Outflow & 0.194 & 0.165 \\
 Optically Thick Compact & 0.061 & 0.104 \\
 \hline
 Sum & 1 & 1  \\
 \hline
 \end{tabular}

\end{table}

\section{Monte Carlo Simulation}\label{sec:sim}
The flux ratio of the two Raman \ion{O}{6} bands $F(6825)/F(7082)$ can be used as a rough proxy to classify a symbiotic star into D-type or S-type. According to \cite{schmid99}, in fact, D-types tend to show $F(6825)/F(7082)\sim 6$, while for S-types $F(6825)/F(7082)\sim 3$. 
The difference in $F(6825)/F(7082)$ seems mainly due to the different \ion{H}{1} column density $N_{HI}$ characterizing the Raman-scattering region in the two types of symbiotic stars, where $N_{HI}$ tends to be larger in S-types because of their smaller binary separation compared to D-type systems. 
Our MIKE data shows $F(6825)/F(7082) \sim 4.5$, which is an intermediate value between S- and D- type symbiotic stars.

In this section, we perform Monte Carlo simulations in order to estimate the
representative value of $N_{HI}$ in Sanduleak's star by reproducing
the observed $F(6825)/F(7082)$. 
We consider a cylindrical neutral scattering region characterized  by a column density $N_{H I}$ measured along the cylinder axis. We place it in front of the giant hemisphere that is facing the white dwarf. 
We align the scattering region, assumed stationary with respect to the white dwarf, so that the cylinder axis coincides with the axis connecting the two stars.
We vary values of $N_{H I}$ from $1\times 10^{22} {\rm\ cm^{-2}}$ to $5\times 10^{23} {\rm\ cm^{-2}}$ when the simulated profile for the Raman 7082~\AA\ band becomes stronger than the observed one. 

The simulation starts with a generation of an \ion{O}{6} photon (according to the input profiles given in Fig.~\ref{input}), which subsequently enters the scattering region. The \ion{O}{6} photon then moves around through Rayleigh scattering processes and escape from the scattering region once Raman-scattering occurs. A more detailed description of this family of Monte Carlo simulations can be found in \cite{heo15}.

\begin{figure*}
\centering
\epsscale{2.}
\plottwo{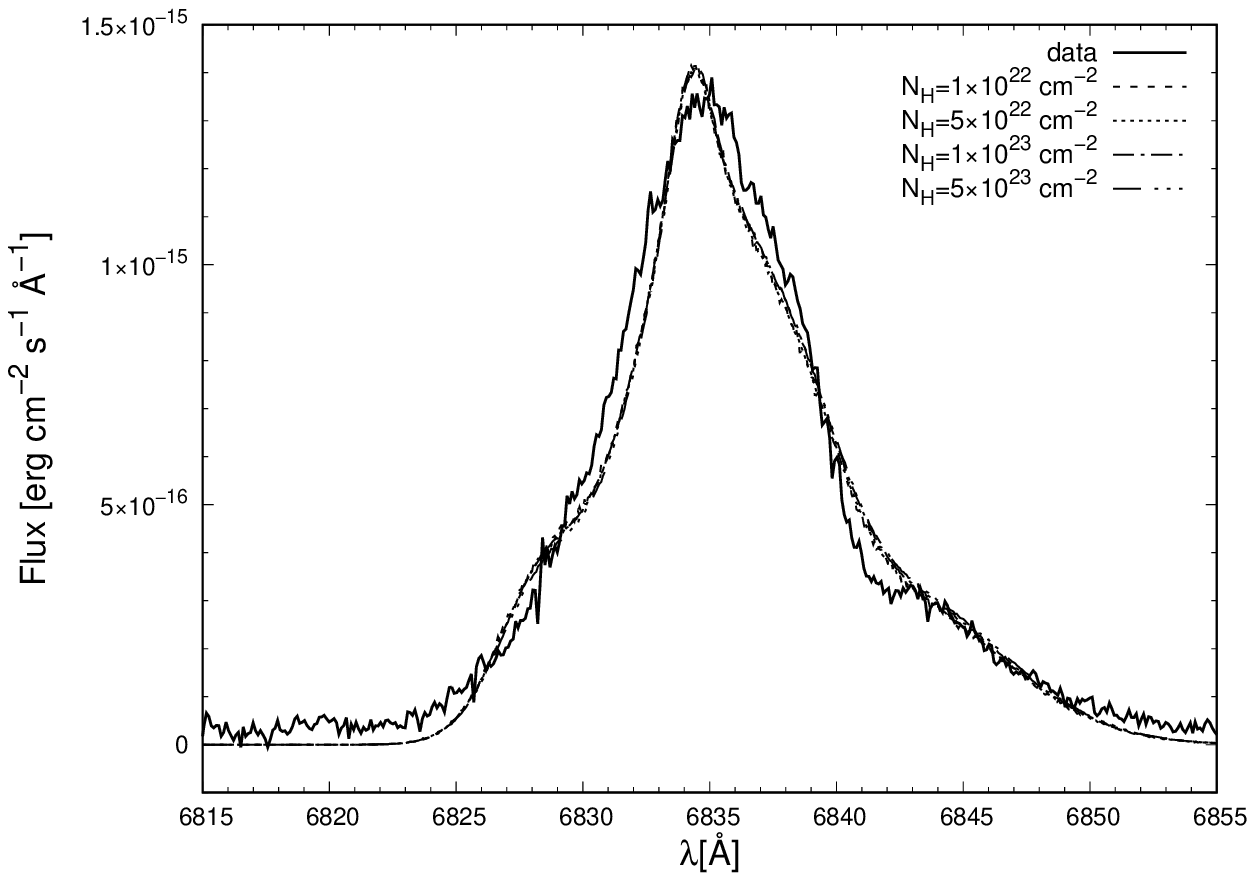}{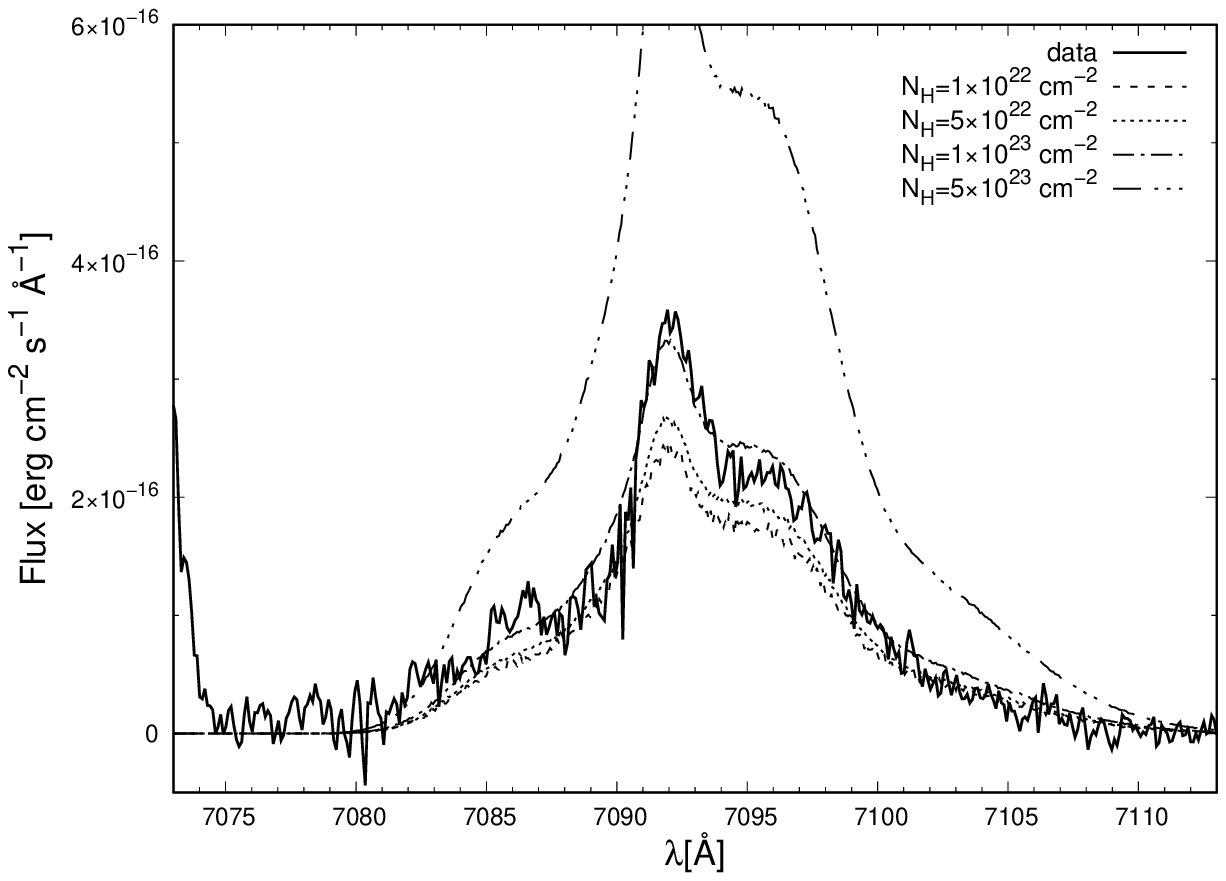}
\caption{Monte Carlo simulations of Raman-scattered \ion{O}{6}~$\lambda\lambda$~1032 (left panel) and 1038 (right panel) for various $N_{HI}$. 
The solid line shows the observation, while the dotted lines represents the results of our Monte Carlo simulations.}
\label{mc_result}
\end{figure*}

In Fig.~\ref{mc_result}, we show the results of our Monte Carlo simulations of the Raman-scattered \ion{O}{6}~$\lambda\lambda$~1032, 1038 profiles for various values of $N_{H I}$, superposed on the 2010 \textit{MIKE} data. 
We normalize the simulated spectra with the observed Raman 6825~\AA\ band and
look for the best fitting profile of the observed Raman 7082~\AA\ band. 
A good fit is obtained for $N_{HI} \sim 1\times 10^{23} {\rm\ cm^{-2}}$. Our Monte
Carlo result shows that $F(6825)/F(7082) >5$ appropriate for D-type symbiotic stars are obtained for $N_{HI} <10^{22}{\rm\ cm^{-2}}$. On the other hand, for $N_{HI}>3\times 10^{23}{\rm\ cm^{-2}}$ the flux ratio becomes lower than 4, resulting in poor fit.


\section{Discussion and Concluding Remarks}\label{sec:dis}

In this work we have presented a systematic analysis of the Raman-scattered \ion{O}{6} bands in Sanduleak's star.
Following a Gaussian decomposition scheme, we successfully fit the profiles of Raman \ion{O}{6}~6825 \AA\ and 7082 \AA\ bands by assuming that the \ion{O}{6} nebula can be decomposed into several emitting components, to be identified with the accretion disk, the bipolar outflow and a further compact, optically thick component.
We have also performed Monte Carlo simulations, from which we propose that the
neutral region of Sanduleak's star is characterized by $N_{HI} \sim 1 \times 10^{23} {\rm\ cm^{-2}}$.

When remembering the very definition of $\Delta V_{O VI}$ and $\Delta V_{atomic}$, it is suggesting to interpret the difference between $\Delta V_{O VI}$ and $\Delta V_{atomic}$ (of the order of $14{\rm\ km\ s^{-1}}$) as the terminal wind velocity of the cool component, that so far has escaped any direct detection and it's at the origin of the controversial status of Sanduleak's star. 

One physical mechanism that may contribute to the redward spectral shift can be found in the asymmetric frequency diffusion of resonantly scattered line photons in an expanding medium. In a static medium, a line photon arising from a permitted transition is usually optically thick at its line center. This implies that the escape of a resonance line photon is made through frequency diffusion after a large number of local scatterings. However, in an expanding medium, a line photon shifted blueward in the frequency space will be scattered after it travels to a spot where the resonance condition is satisfied. In this way, the blue part of line radiation is significantly suppressed to form a P~Cyg profile. Naively, we expect that resonance line radiation emergent locally in an emission region with an expansion speed of $\Delta V$ is shifted redward by the same quantity \citep[e.g.,][]{schmid99}.
In a complex emission region consisting of a number of emission components, each redward
shift is convolved in a complicated way that unfortunately can not be disentangled by our present observations.

Spectropolarimetry would have the power of shedding much more light on the detailed structure 
of the \ion{O}{6} emission region in symbiotic stars. Spectropolarimetric observations performed by \cite{harries96}
showed for example that Raman-scattered \ion{O}{6} bands are strongly polarized, and that in many cases
the red wing is polarized in the direction perpendicular to the polarization direction 
of the main part. 

Interestingly, this behavior is naturally expected in our decomposition scheme 
in which the band red wing originates from the bipolar outflow, that indeed moves away
in the direction perpendicular to the accretion disk. Our profile decomposition thus foresees that the emitting component associated with the outflow will be polarized in the direction perpendicular to the jet axis, whereas the
remaining part will be polarized along it. Because the bipolar outflow component is mainly dominant in the red wing,
we also expect the degree of linear polarization to be
significant there. In a similar way, the blue wing will exhibit a fairly high degree of
polarization along the jet axis. However, in the red main part where the REP component and the bipolar outflow
component contribute, compensation of the two oppositely polarized components
may result in a lower degree of polarization, strongly dependent on the relative contribution of each component to the overall profile.

In this context, spectropolarimetric observations represent a crucial validation test of our decomposition scheme and at the same time a unique opportunity to clarify the astrophysical nature of Sanduleak's star.

\acknowledgments

We are grateful to the anonymous referee for useful comments. 
We also thank Kwang-il Seon for his help. 
\textit{FUSE} data presented in this paper were obtained from the Mikulski Archive for Space Telescopes (MAST). STScI is operated by the Association of Universities for Research in Astronomy, Inc., under NASA contract NAS5-26555. Support for MAST for non-HST data is provided by the NASA Office of Space Science via grant NNX09AF08G and by other grants and contracts.
This research was supported by the Korea Astronomy and Space Science Institute
under the R\&D program (Project No. 2015-1-320-18) supervised by the Ministry of Science, ICT and Future Planning.


\begin{thebibliography}{}

\bibitem[\protect\citeauthoryear{Allen}{1980}]{allen80}
Allen, D.~A. 1980, ApL, 20, 131

\bibitem[\protect\citeauthoryear{Allen}{1987}]{allen87}
------. 1987, Forbidden Emission Lines - II, Springer Netherlands, doi:10.1007/978-94-009-3769-7\_59

\bibitem[\protect\citeauthoryear{Angeloni et al.}{2011}]{angeloni11}
Angeloni, R., Di Mille, F., Bland-Hawthorn, J., \& Osip, D. J. 2011, \apj, 743, L8

\bibitem[\protect\citeauthoryear{Belczy\'nski et al.}{2000}]{belc00}
Belczy\'nski, K., Miko\l ajewska, J., Munari, U., Ivison, R.~J., \& Friedjung, M. 2000, \aaps, 146, 407

\bibitem[\protect\citeauthoryear{Bernstein et al.}{2003}]{bernstein03}
Bernstein, R., Shectman, S.~A., Gunnels, S.~M., Mochnacki, S., \& Athey, A.~E.  2003, Proc. SPIE, 4841, 1694


\bibitem[\protect\citeauthoryear{Birriel et al.}{1998}]{birriel98}
Birriel J., Espey B.~R., \& Schulte-Ladbeck R. E. 1998, \apj, 507, L75

\bibitem[\protect\citeauthoryear{Birriel et al.}{2000}]{birriel00}
------. 2000, \apj, 545, 1020

\bibitem[\protect\citeauthoryear{Ciddor}{1996}]{ciddor96}
Ciddor, P.~E. 1996, ApOpt, 35, 1566

\bibitem[\protect\citeauthoryear{Crowley et al.}{2008}]{crowley08}
Crowley, C., Espey, B. R., \& McCandliss, S. R. 2008, \apj, 675, 711

\bibitem[\protect\citeauthoryear{de Val-Borro et al.}{2009}]{deval09}
de Val-Borro, M., Karovska, M., \& Sasselov, D. 2009, \apj, 700, 1148

\bibitem[\protect\citeauthoryear{Dixon et al.}{2007}]{dixon07} 
Dixon, W.~V., Sahnow, D.~J., Barrett, P.~E., et al. 2007, PASP, 119, 527

\bibitem[\protect\citeauthoryear{Eyres et al.}{1995}]{eyres95} 
Eyres, S.~P.~S., Kenny, H.~T., Cohen, R.~J. et al. 2007, PASP, 119, 527

\bibitem[\protect\citeauthoryear{Harries \& Howarth}{1996}]{harries96}
Harries, T.~J., \& Howarth, I.~D. 1996, \aaps, 119, 61

\bibitem[\protect\citeauthoryear{Heo \& Lee}{2015}]{heo15}
Heo, J.-E., \& Lee, H.-W. 2015, J. Korean Astron. Soc., 48, 105

\bibitem[\protect\citeauthoryear{Huarte-Espinosa et al.}{2013}]{huarte13}
Huarte-Espinosa, M., Carroll-Nellenback, J., Nordhaus, J., Frank, A., \& Blackman, E. G. 2013, \mnras,  433, 295

\bibitem[\protect\citeauthoryear{Joy \& Swings}{1945}]{joy45}
Joy, A.~H., \& Swings, P. 1945, \apj, 102, 353 

\bibitem[\protect\citeauthoryear{Kang \& Lee}{2008}]{kang08}
Kang, E.-H., \& Lee, H.-W. 2008, J. Korean Astron. Soc., 41, 49

\bibitem[\protect\citeauthoryear{Lee \& Kang}{2007}]{lee07}
Lee, H.-W., \& Kang, S. 2007, \apj, 669, 1156

\bibitem[\protect\citeauthoryear{Luna \& Costa}{2005}]{luna05}
Luna, G. J. M., \& Costa, R. D. D. 2013, \aap, 435, 1087

\bibitem[\protect\citeauthoryear{Mastrodemos \& Morris}{1998}]{mastro98}
Mastrodemos, N., \& Morris, M. 1998, \apj, 497, 303

\bibitem[\protect\citeauthoryear{Miko\l ajewska et al.}{2006}]{miko06}
Miko\l ajewska, M., Friedjung, M., \& Quiroga, C. 2006, \aap, 460, 191


\bibitem[\protect\citeauthoryear{Moore}{1979}]{moore79}
Moore, C.~E. 1979, NSRDS-NBS 3, \S 8

\bibitem[\protect\citeauthoryear{Moos et al.}{2000}]{moos00} 
Moos, H.~W., Cash, W.~C., Cowie, L.~L., et al.\ 2000, \apj, 538, L1 

\bibitem[\protect\citeauthoryear{Munari}{2012}]{munari12} 
Munari, U.\ 2012, JAAVSO, 40, 572

\bibitem[\protect\citeauthoryear{Munari \& Zwitter}{2002}]{munari02}
Munari, U., Zwitter, T. 2002, \aap, 383, 188 

\bibitem[\protect\citeauthoryear{Nichols \& Slavin}{2009}]{nichols09}
Nichols, J., \& Slavin, J.~D. 2009, \apj, 669, 902

\bibitem[\protect\citeauthoryear{Nussbaumer et al.}{1989}]{nussbaumer89}
Nussbaumer, H. Schmid, H.~M., \& Vogel, M. \aap, 211, L27

\bibitem[\protect\citeauthoryear{Sahai et al.}{2011}]{sahai11}
Sahai, R., Morris, M.~R., \& Villar, G.~G. 2011, AJ, 141, 134
  
\bibitem[\protect\citeauthoryear{Sahnow et al.}{2000}]{sahnow00} 
Sahnow, D.~J., Moos, H.~W., Ake, T.~B., et al.\ 2000, \apj, 538, L7 
  
\bibitem[\protect\citeauthoryear{Sanduleak}{1977}]{sanduleak77}
Sanduleak, N., 1977, IBVS, 1304, 1

\bibitem[\protect\citeauthoryear{Schild \& Schmid}{1996}]{schild96}
Schild, H., \& Schmid, H.~M. 1996, \aap, 310, 211

\bibitem[\protect\citeauthoryear{Schmid}{1989}]{schmid89}
Schmid, H.~M. 1989, \aap, 211, L31

\bibitem[\protect\citeauthoryear{Schmid et al.}{1999}]{schmid99}
Schmid, H.~M., et al. 1999, \aap, 348, 950

\bibitem[\protect\citeauthoryear{Schmid et al.}{2000}]{schmid00}
Schmid, H.~M., R. Corradi, J. Krautter, \& H. Schild, 2000, \aap, 355, 261

\bibitem[\protect\citeauthoryear{Shagatova et al.}{2016}]{shagatova16}
Shagatova, N., Skopoal, A., \& Carikov\'a, Z. 2016, \aap, 588, 83

\bibitem[\protect\citeauthoryear{Solf}{1984}]{solf84}
Solf, J. 1984, \aap, 139, 296

\bibitem[\protect\citeauthoryear{Walder et al.}{2008}]{walder08}
Walder, R., Folini, D., \& Shore, S.~N. 2008, \aap, 484, L9

\bibitem[\protect\citeauthoryear{Wood et al.}{2002}]{wood02}
Wood et al. 2002, \apjs, 140, 91

\end{thebibliography}
\end{document}